\documentclass[twocolumn,traditabstract, printer]{aa}
\usepackage[varg]{txfonts}

\usepackage{graphicx}
\usepackage{natbib}
\usepackage{color}
\usepackage{subfig}
\usepackage[switch]{lineno} 
\usepackage{xspace}
\usepackage{microtype}
\usepackage{amssymb}
\def\link_col{blue}
\usepackage{xspace}
\usepackage{url}
\usepackage{wasysym}
\usepackage{caption}

\def\fermi{{\it Fermi}-LAT \xspace}
\def\planck{{\it Planck}\xspace}

\def\grays{$\gamma$-rays\xspace}
\def\gray{$\gamma$-ray\xspace}

\begin{document}

\title{Investigation of cosmic ray penetration with wavelet cross-correlation analysis}
\author{Rui-zhi Yang}
\institute{Max-Planck-Institut f{\"u}r Kernphysik, P.O. Box 103980, 69029 Heidelberg, Germany.}
\date{Received:  / Accepted: } 

\abstract {}
{
We use  \fermi and \planck data to calculate  the cross correlation between \gray signal and gas distribution in different scales in giant molecular clouds (GMC). Then we investigate the cosmic rays (CRs) penetration in GMCs with these informations. }
{
We use the wavelet technique to decompose both the \gray and dust opacity maps in different scales, then we calculate the wavelet cross correlation  functions in these scales. We also define   wavelet response as an analog to the impulsive response in Fourier transform and calculate that in different scales down to \fermi angular resolution. 
}
{ The  \gray maps above 2 GeV show strong correlation with the dust opacity maps, the correlation coefficient is larger than 0.9 above a scale of 0.4 degree.The derived wavelet response is uniform in different scales. 
}
{
We argue that the CR above 10 GeV can penetrate the giant molecular cloud freely and the CRs distributions in the same energy range  are uniform down to parsec scale. 
}

\keywords{Gamma rays: ISM -- (ISM:) cosmic rays }
\maketitle
\section{Introduction}
Cosmic rays (CRs) play an important role in the determination of ionization level in molecular clouds (MC), and thus have an impact on the dynamical process and star formation \citep[for a review, see][]{dalgarno06}. The penetration of CRs inside the MCs is a key issue to evaluate the CR density therein. On the other hand, MCs are also regarded as the CR barometers \citep{aharonian91,aharonian01,casanova10} and thus are important targets in the \gray astronomy.  The exclusion of CRs from MCs may reduce the \gray brightness and alter the spectral shape.  The issue of the penetration or exclusion of CRs from MCs has been investigated by several groups\citep{skilling76, dogiel90, gabici07, morlino15}. However, in these papers quite different results have been derived, from almost free-penetration to exclusion of CRs up to $100$ GeV.  Generally, the propagation of CRs depends strongly on the turbulence spectrum inside MCs. The high gas density in MCs cause high damping rate, which reveal that the propagation of CRs inside MCs are dominated by convection rather than diffusion. In such a scenario the CRs can freely penetrate into the dense core of the MCs except the low energy particles, which may suffer from the ionization losses \citep{morlino15}.  On the other hand \citet{istomin13} argue that the magnetic turbulence can also be generated in the weak ionized gas by the kinematic turbulence in the neutral gas. Thus the CRs exclusion can occur due to the slower diffusion induced by the higher turbulence level \citep[see e.g.][]{gabici07}.

Various observational efforts have also been made on  the interaction of CRs and MCs, both on \grays \citep[see e.g. ][]{fermiorion, fermimc,yang14,yang15, planck_cha} and ionizations \citep{padovani09, nath12}. However, the \gray analysis treated MCs as a whole rather than investigated the CRs penetration in different scales, while the ionization study can only account for low energy part of CRs distributions  and the contamination from electrons is hard to  exclude. Direct observation evidence is still lacking in determination of the different scenarios mentioned above.  In this paper we analyze the \fermi  and \planck dust opacity data in Taurus and Orion A region and use a wavelet cross-correlation technique \citep[see e.g.][]{frick01,arshakian16}  to study correlation between \gray maps and dust opacity maps in different scales down to the instrument angular resolution. The correlation factor and wavelet response derived can give us unambiguous information on the CRs density in different scales.  We also compare the derived cross correlation and  wavelet response with the simulation results to test the technique. 

This paper is organized as follows. In Sec. 2 we describe the data reduction and maps used in this study, in Sec. 3 we briefly describe the wavelet decomposition technique and calculate the wavelet cross correlation and wavelet impulsive response at different scales,  in Sec. 4 we perform simulations with different scenarios and compare them with the results from the data,  and in Sec. 5 we conclude and discuss the implication of our results.

\section{\gray and dust column maps}
We chose  Orion A and Taurus region as our targets. Both of them are famous nearby giant molecular clouds and significantly detected in \grays \citep{yang14}. Orion A is located 500 pc away from the Sun and has a mass of about $1.2\times 10^5 ~\rm M_{\odot}$, while Taurus is 140 pc away  and has a mass of about $0.3\times 10^4 ~\rm M_{\odot}$.  The diffuse emission induced  mainly by neutral pion decay in the interaction of CRs and ambient gas dominates the contributions from the point sources in both regions, which makes it a ideal place to investigate the interaction of CRs and ambient gas.

For \fermi data we have selected observations with over the period which covers seven years exposure time (MET 239557417 -- MET 451533077).
For the data reduction, we used the standard LAT analysis software package \emph{v9r33p0}\footnote{\url{http://fermi.gsfc.nasa.gov/ssc}}. 
We want to use the data with high angular resolution to investigate the very inner core of MCs.   The \fermi angular resolution improves with the rising energy. The 68\% containment angles are about $5.0^{\circ}, 0.8^{\circ}, 0.5^{\circ}$, and $0.2^{\circ}$ at 0.1, 1, 2 and 10 GeV, respectively \footnote{\url{https://www.slac.stanford.edu/exp/glast/groups/canda/lat\_Performance.htm}}. Thus the higher angular resolution can be only achieved at higher energy bands, where the statistics are much worse.  In wavelet study the shooting noise may dominate the signal in small scales with lower significance.  On the other hand,  the front converted photon events have a better angular resolution, the 68\% containment angle at 2 GeV is about $0.35^{\circ}$, rather than $0.5^{\circ}$ degree for all events.  Thus we selected front converted events with energies exceeding 2000~MeV to satisfy the requirement on both statistics and angular resolution. 

\begin{figure*}[htb]
\centering
\includegraphics[width=0.4\textwidth,angle=0]{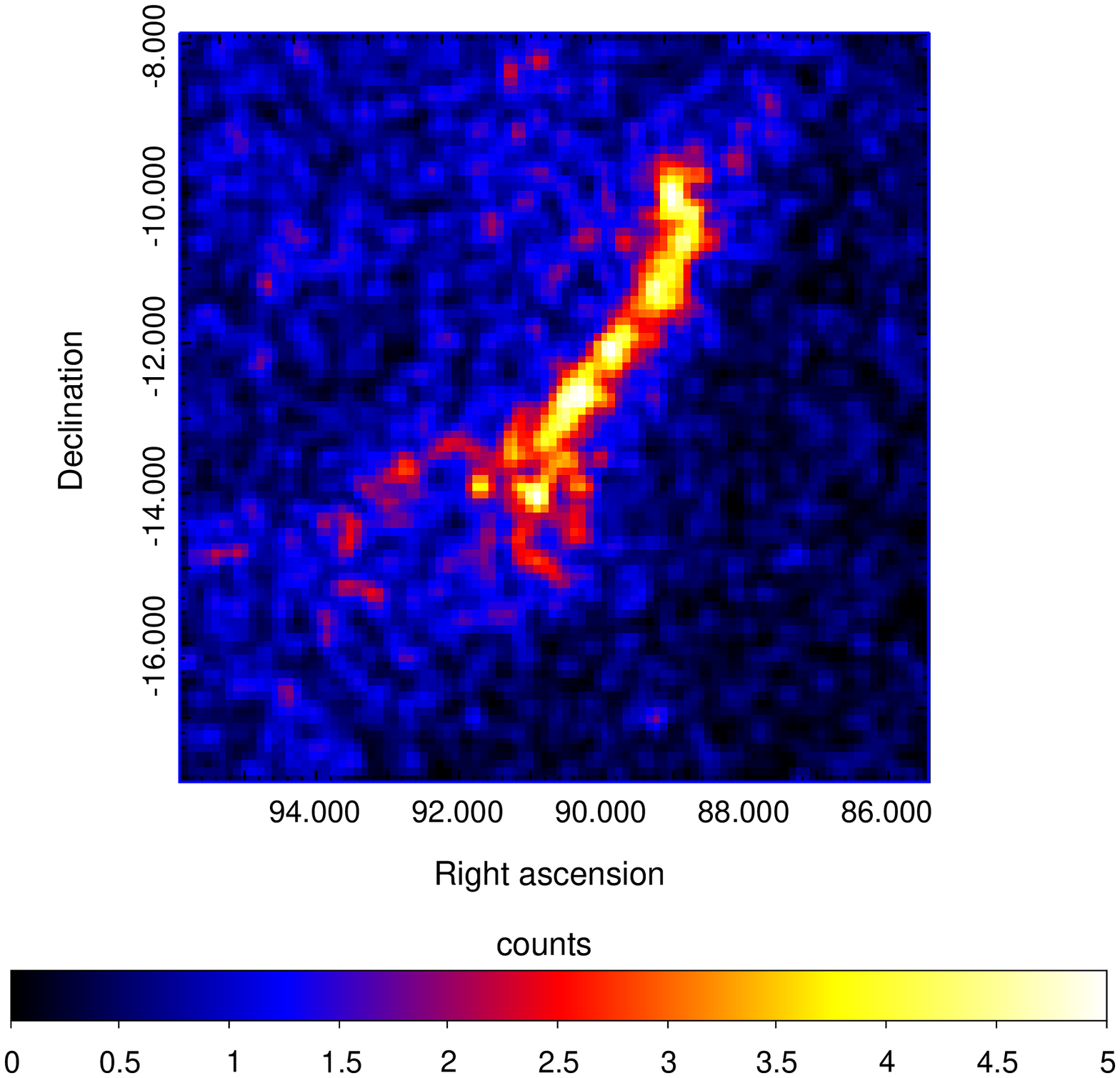}
\includegraphics[width=0.4\textwidth,angle=0]{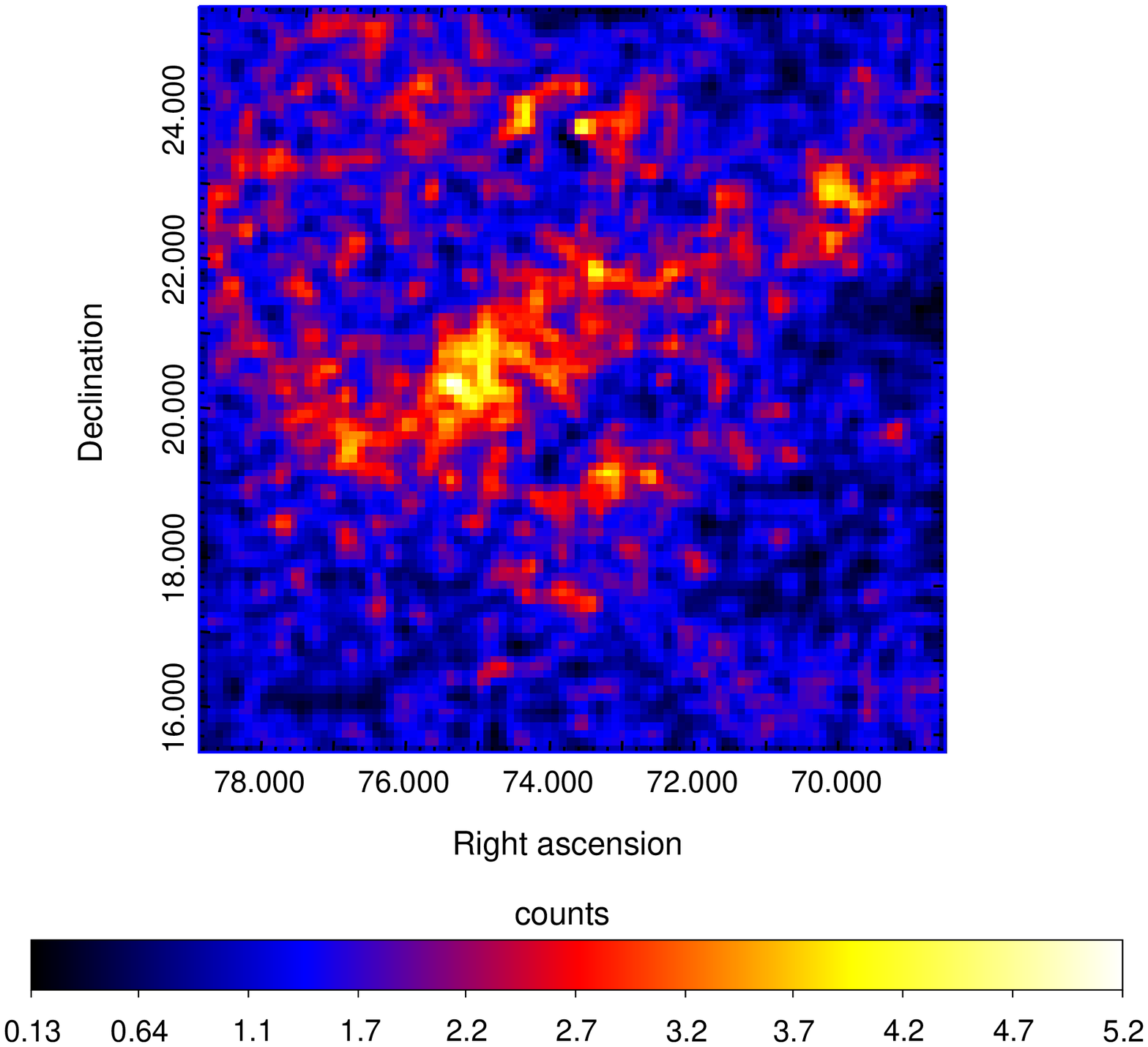}\\
\includegraphics[width=0.4\textwidth,angle=0]{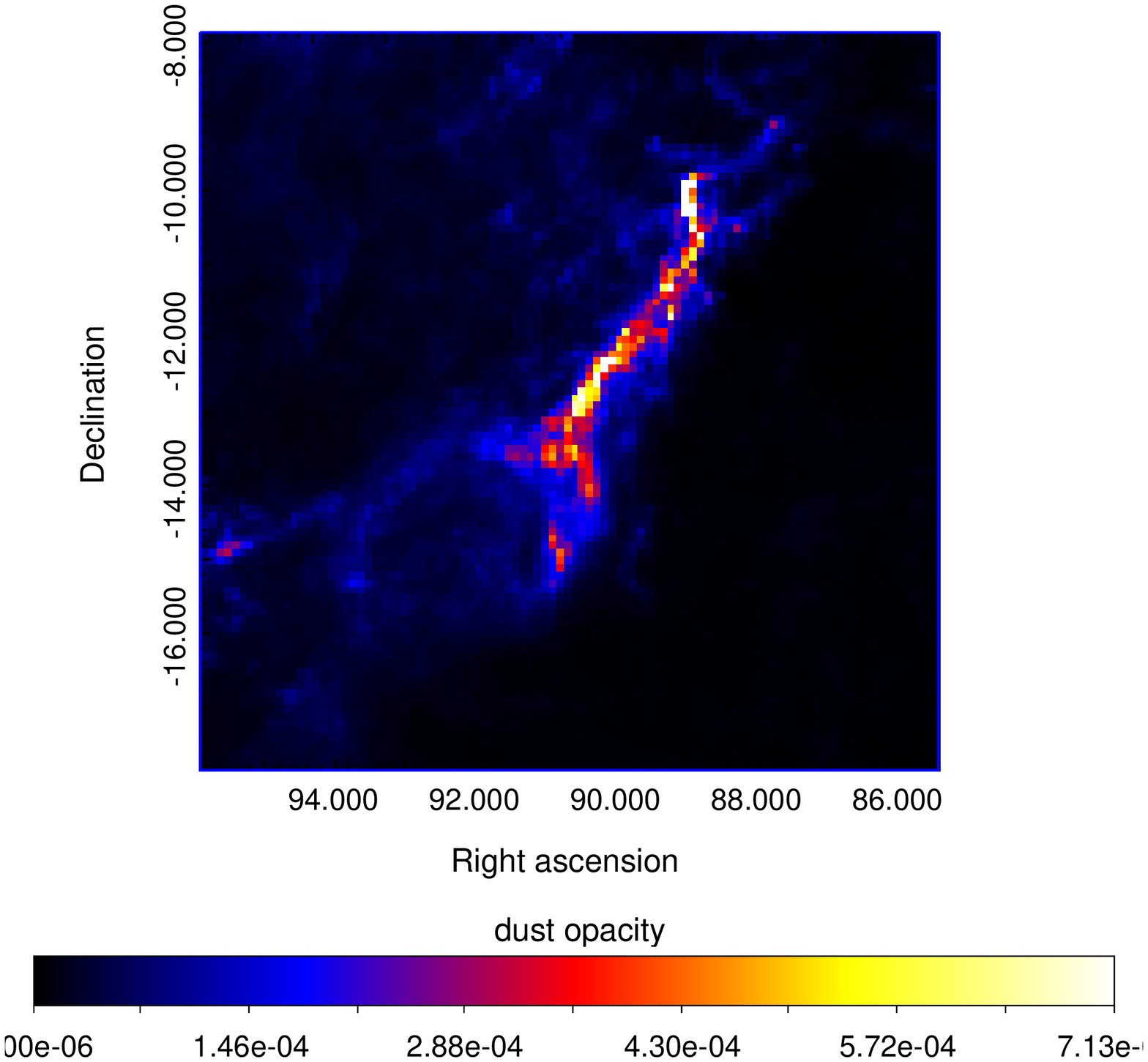}
\includegraphics[width=0.4\textwidth,angle=0]{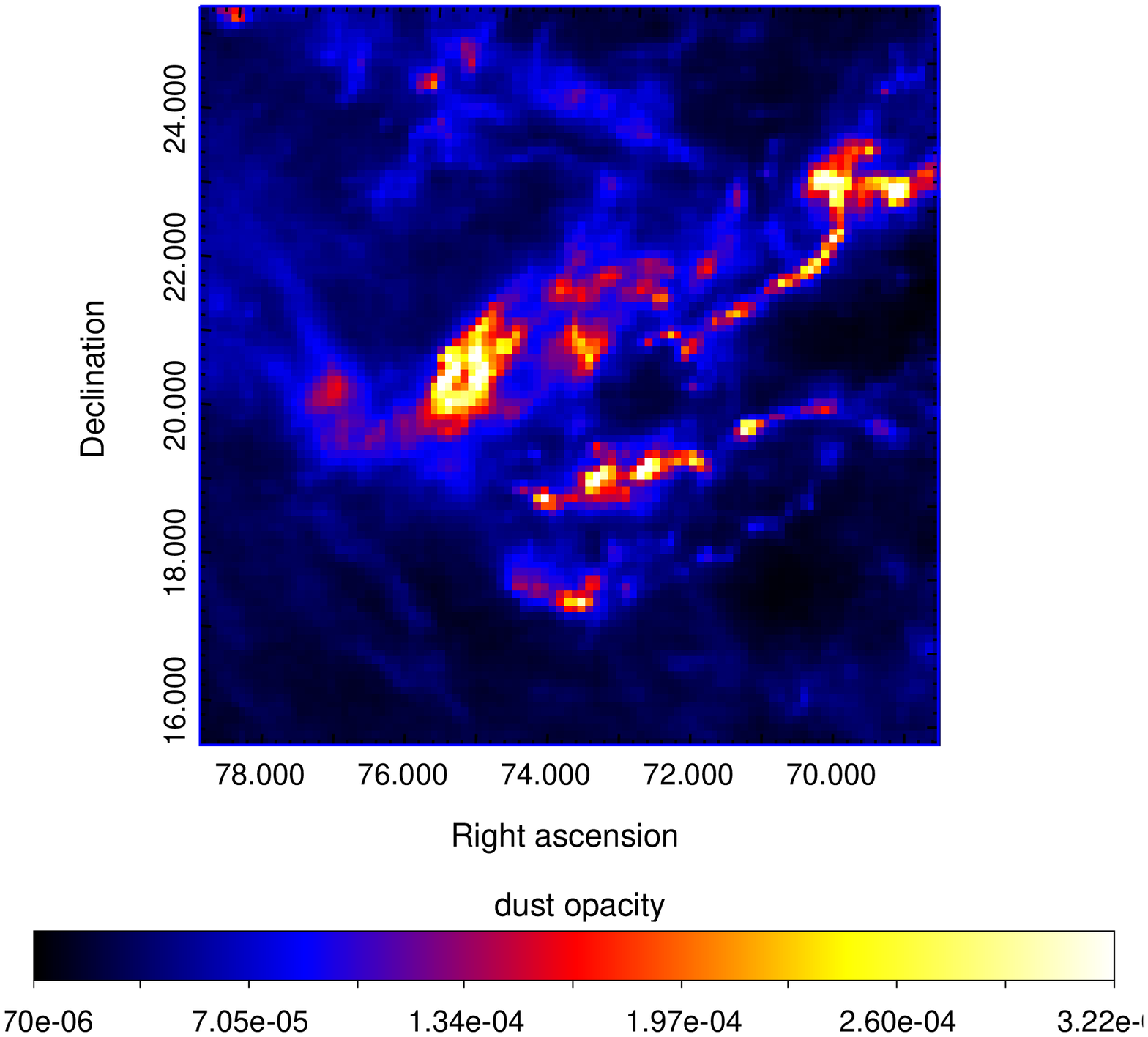}\\
\caption{{\it Upper panel}:  \gray counts map above 2 GeV after point source subtraction in the region Orion A (left) and Taurus (right). {\it Lower panel}: $\tau_{353}$ maps for Orion A (left) and Taurus (right)  }
\label{fig:1}
\end{figure*}

The region-of-interest (ROI) was defined to be a $10^ \circ \times 10^ \circ$ square centered on the position of Taurus and Orion A, respectively. 
In order to reduce the effect of the background related to the Earth's albedo, we excluded from the analysis the time intervals when the Earth was in the field-of-view (more specifically,  when the centre of the field-of-view was  $52^ \circ$ above the zenith), as well as the time intervals when parts of the ROI were observed at zenith angles $> 100^ \circ$. 
Although point sources only contribute small part of the total emission in this regions, to minimize these contaminations we subtract the best fit point sources' flux in both regions. The P8R2\_v6 version of the post-launch instrument response functions (IRFs) was adopted in the likelihood fitting. We also use the 3FGL catalog \citep{3fgl}  and standard diffuse emission templates \footnote{gll\_iem\_v06.fit and iso\_P8R2\_SOURCE\_V6\_v06.txt, available at \url{http://fermi.gsfc.nasa.gov/ssc/data/access/lat/BackgroundModels.html}}.   The spectral parameter of the point sources with ROI and the diffuse components are left free in the fitting.  Then we subtract the best fitting point sources model maps generated by {\it gtmodel} tool from the raw counts map. The point sources subtracted counts maps can be found in Fig.\ref{fig:1}. 

The \planck dust opacity maps are used to model the gas distribution. The detailed description of the dust opacity maps can be found in \citet{planck}. Although \citet{planck2013-11} argue that the radiance may be a more proper tracer of gas in the diffuse interstellar medium, we still chose $\tau_{353}$ as the tracer since the column in Taurus and Orion A is relatively high ($\sim 10^{22}~\rm cm^{-2}$). The $\tau_{353}$  map can also be found in Fig.\ref{fig:1}. And we use the relation  
\begin{equation}
N_H =   \tau_m(\lambda)\left[\left(\frac{\tau_D(\lambda)}{N_H}\right)^{dust}\right]^{-1},
\end{equation} 
where $(\tau_D/N_H)^{dust}_{353{\rm~GHz}}=1.18\pm0.17\times10^{-26}$~cm$^2$ \citep{planck} to convert dust opacity into gas column.  

\section{wavelet transform and wavelet cross correlation}
Wavelet transform of function $f({\bf x})$ is defined as 
\begin{equation}
W(a,{\bf x})=\frac{1}{a^{\kappa}}\int \int f({\bf x'}) \psi^*(\frac{{\bf x-x'}}{a})d{\bf x'},
\end{equation}
Here ${\bf x}$ is the physical coordinates, $\psi({\bf x})$ is the
analysing wavelet (real or complex, * indicates the complex
conjugation), $a$ is the scale parameter, and $\kappa$ is a normalization
parameter. 

\begin{figure*}[htb]
\centering
\includegraphics[width=0.4\textwidth,angle=0]{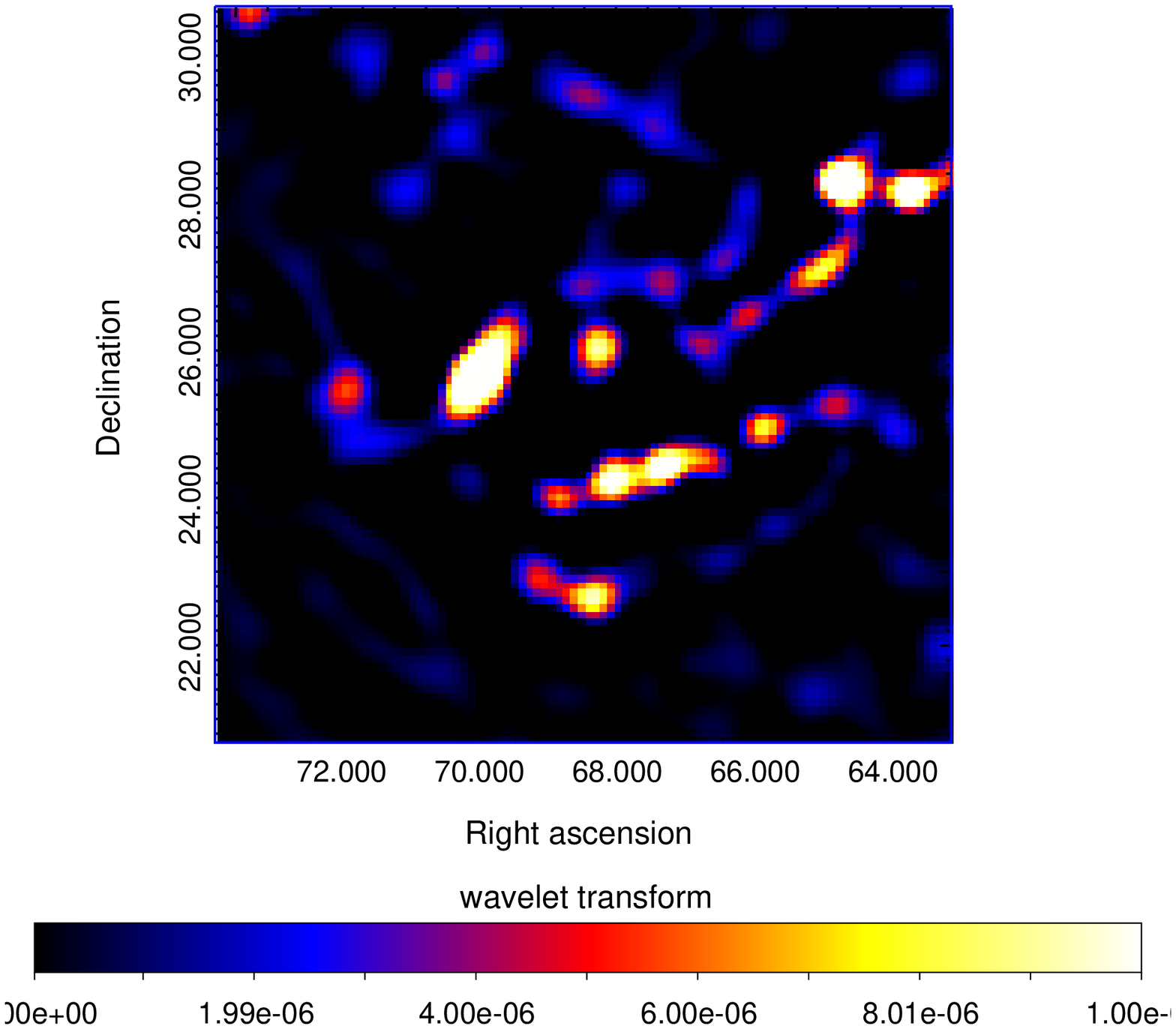}
\includegraphics[width=0.4\textwidth,angle=0]{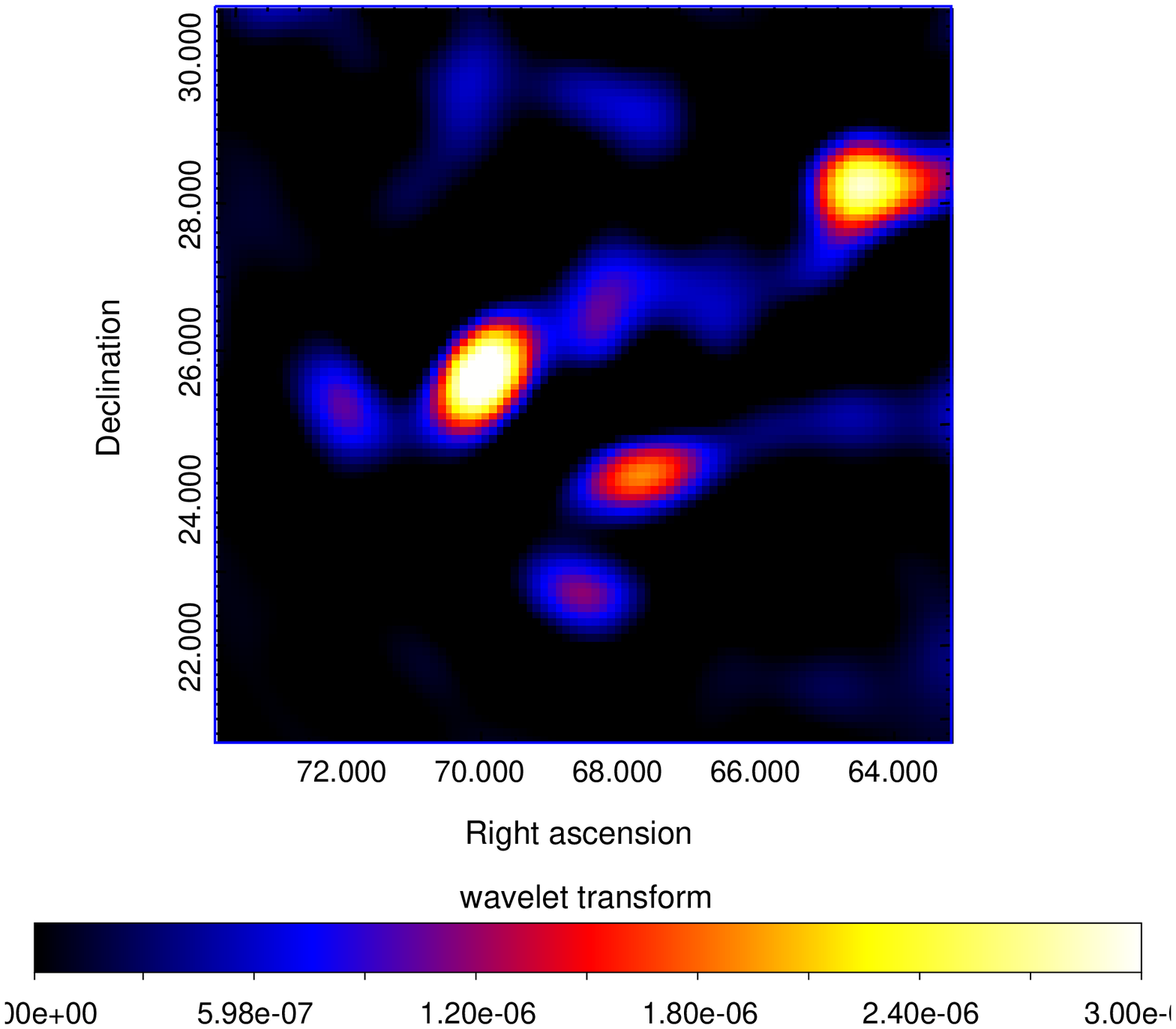}\\
\includegraphics[width=0.4\textwidth,angle=0]{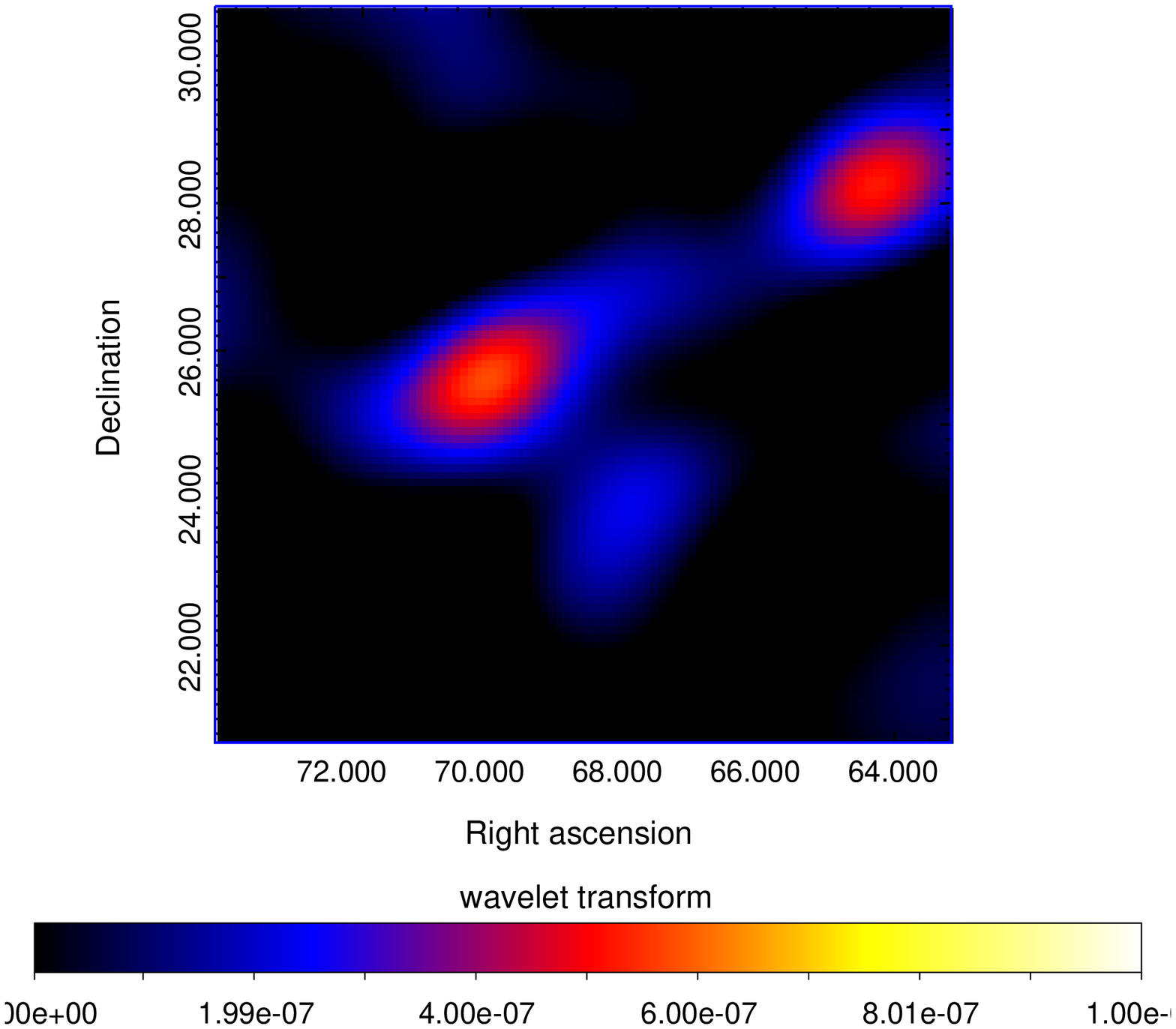}
\includegraphics[width=0.4\textwidth,angle=0]{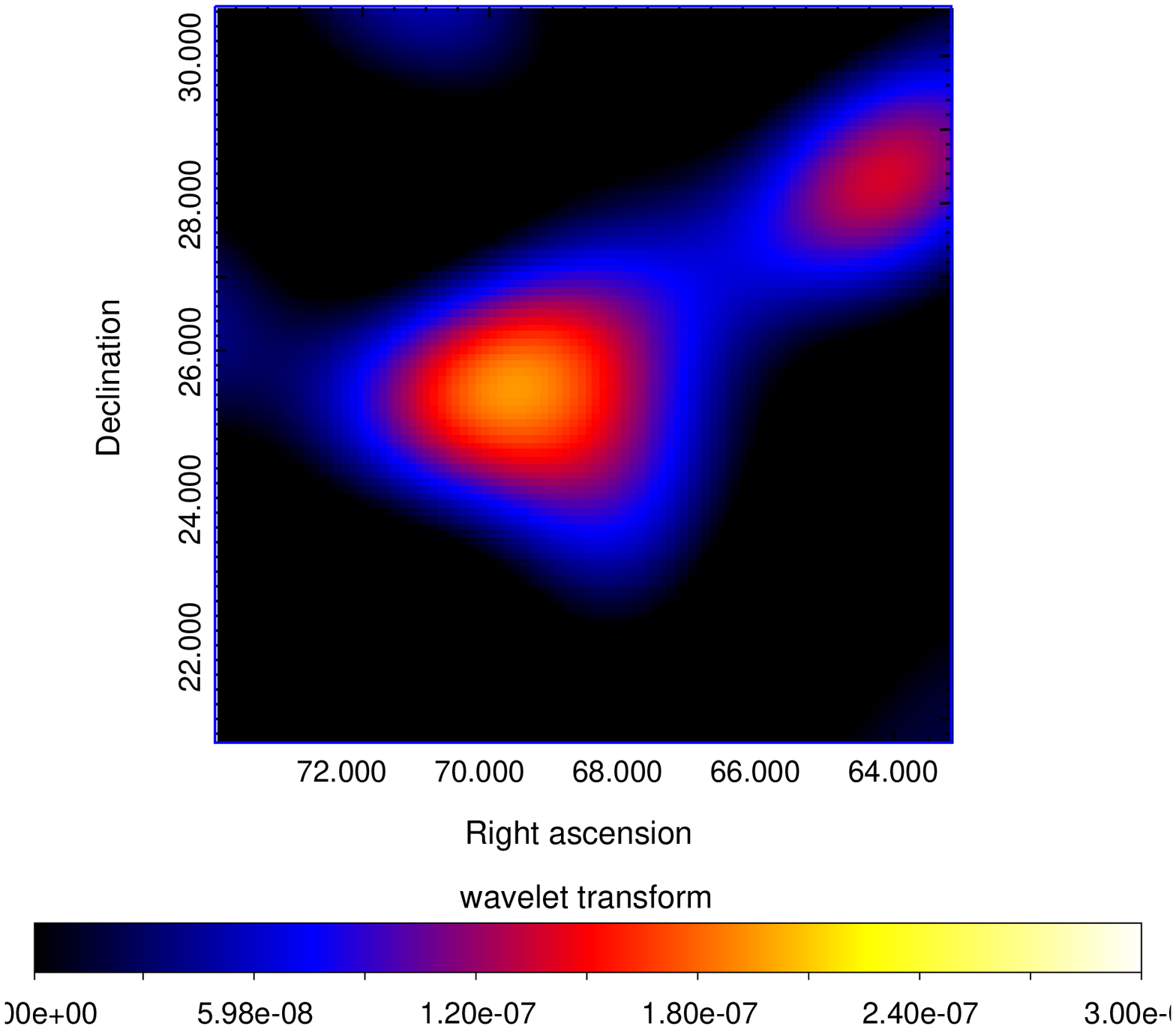}
\caption{Wavelet decomposition of $\tau_{353}$ map of Taurus with the scale of  0.2, 0.5, 1.0, 1.5 degrees (from  top left to bottom right).  }
\label{fig:2}
\end{figure*}

Wavelet transform can decompose images into different scales \citep{frick01}.  Here we use the Mexican hat wavelet transform to decompose the images, which can be expressed as 
\begin{equation}
\psi({\rho})=(2 - \rho^2) e^{-\rho^2/2}.
\end{equation}
 As an example we show the decomposition of $\tau_{353}$ map of Taurus in different scales in Fig.\ref{fig:2}. It is clear that on high resolution maps only the dense cores emerge and in low resolution ones there are only large scale structures.  

The wavelet cross correlation coefficient  of two maps can be defined as 
\begin{equation}
r_w(a)=\frac{\int \int W_1(a,{\bf x}) W^*_2(a,{\bf x})d{\bf x}}{(M_1(a) M_2(a))^{1/2}},
\end{equation}
where 
\begin{equation}
M(a)=\int \int |W(a,{\bf x})|^2 d{\bf x}
\end{equation}
is the wavelet spectrum. 

To calculate the cross correlation coefficients between the \gray maps and dust opacity maps we first smooth both map into the same angular resolution. This is done by following the relation that  $\sigma^2_{f}=\sigma^2_{s}+\sigma^2_{o}$, where $\sigma_{f}$ and $\sigma_{o}$ are the final and origin map resolution and $\sigma_{s}$ is the width of the smoothing kernel. The angular resolution for \fermi  and beam width for Planck maps can be found in the calibration database (CALDB) files in  {\it Fermi Science tools} and \citet{planck15_7}, respectively. If the point spread function has a gaussian form, the 68\% containment angle is identical to $\sigma$ of the gaussian function by definition.  Although the PSF of \fermi is not a perfectly gaussian \footnote{\url{https://www.slac.stanford.edu/exp/glast/groups/canda/lat\_Performance.htm}}, the tail is not relevant for the analysis described here. Thus we assume for \fermi maps $\sigma_{o} = 0.35^{\circ}$. To make full use of the \fermi angular resolution we chose $\sigma_{f}$ to be $0.4^{\circ}$ in this work. 

\begin{figure*}[htb]
\centering
\includegraphics[width=0.4\textwidth,angle=0]{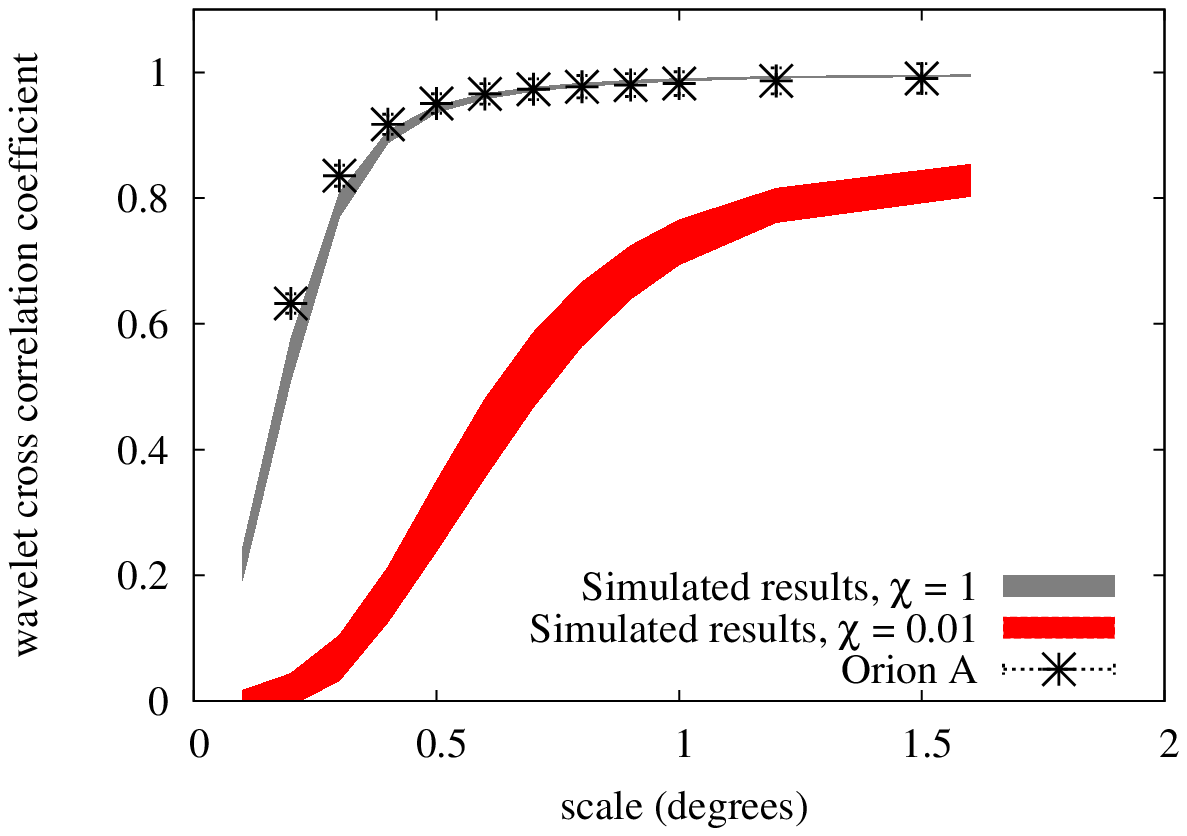}
\includegraphics[width=0.4\textwidth,angle=0]{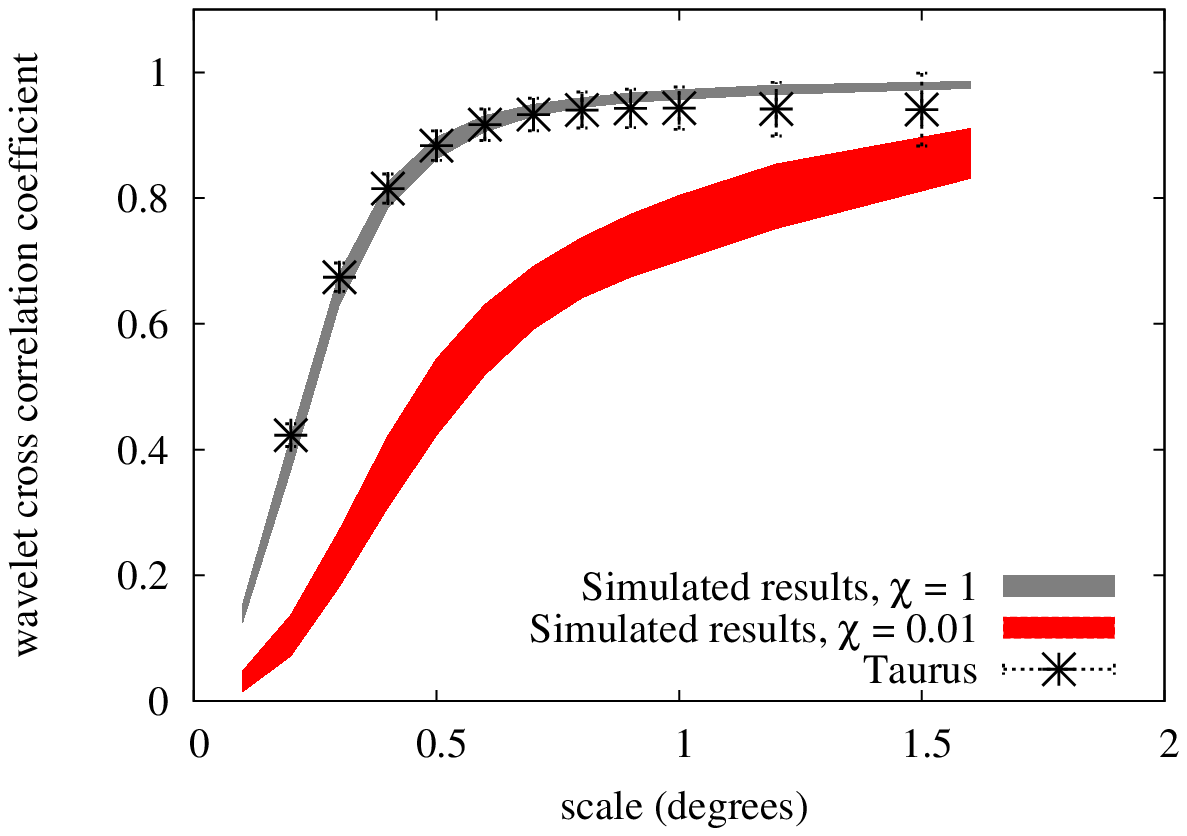}\\

\caption{Wavelet cross-correlation coefficient of Orion A  (left panel ) and Taurus (right panel), respectively. The gray shaded areas show the simulation results assuming a uniform CR density at all scales, while the red areas show the results when assuming a smaller diffusion coefficient inside MC. $\chi$ is defined in Eq. 7.}
\label{fig:3}
\end{figure*}

The results of the cross-correlation coefficients  are shown in Fig.\ref{fig:3}.  The results show a strong correlation above 0.4 degrees (correlation coefficient larger than 0.9).  The error bars are derived using jackknife method.  In jackknife we divided the ROI into 100 groups, $1^{\circ}\times1^{\circ}$ each, then we masked one of the subgroups each time and calculated the corresponding wavelet cross correlation coefficients in the masked map. The final errors can be estimated as $\sqrt{100-1} \sigma_{cc}$, where $\sigma_{cc}$ is the standard deviation of the wavelet cross-correlation coefficients in all the jackknife samples.

\begin{figure*}[htb]
\centering

\includegraphics[width=0.4\textwidth,angle=0]{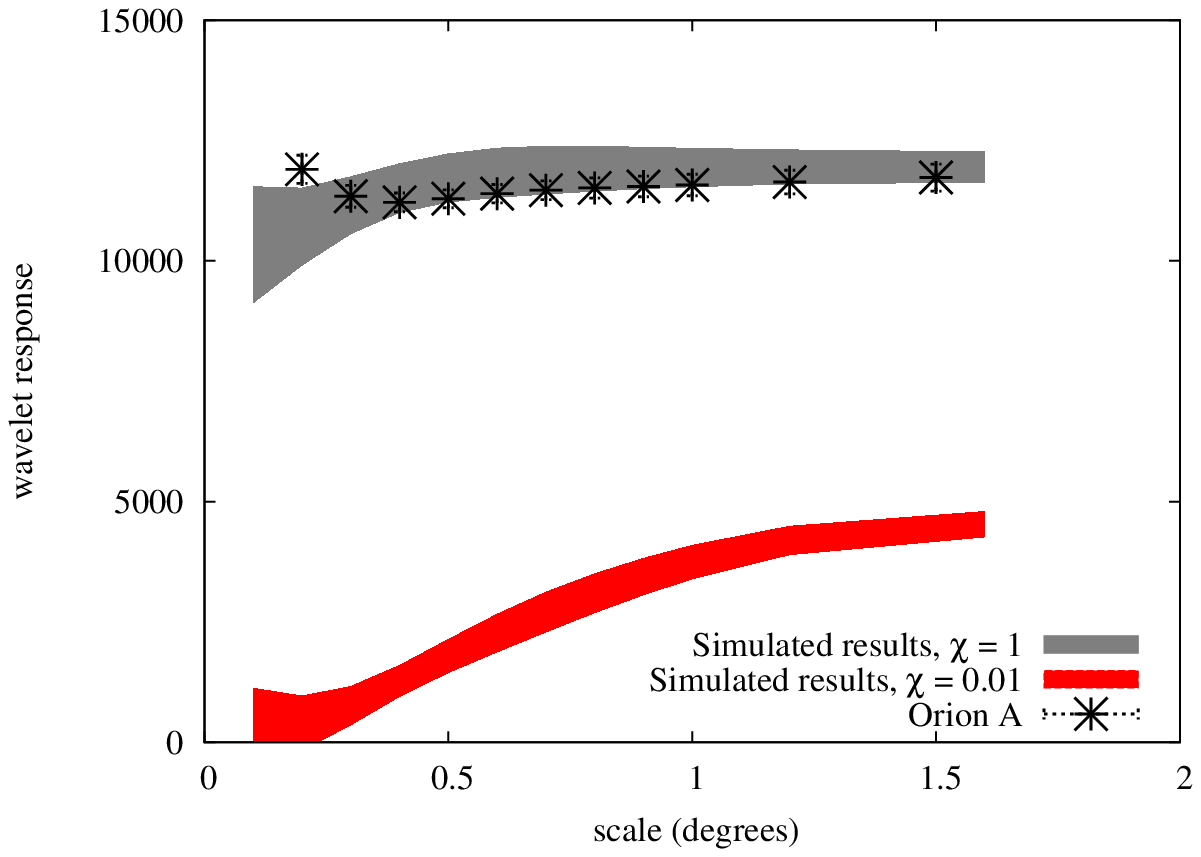}
\includegraphics[width=0.4\textwidth,angle=0]{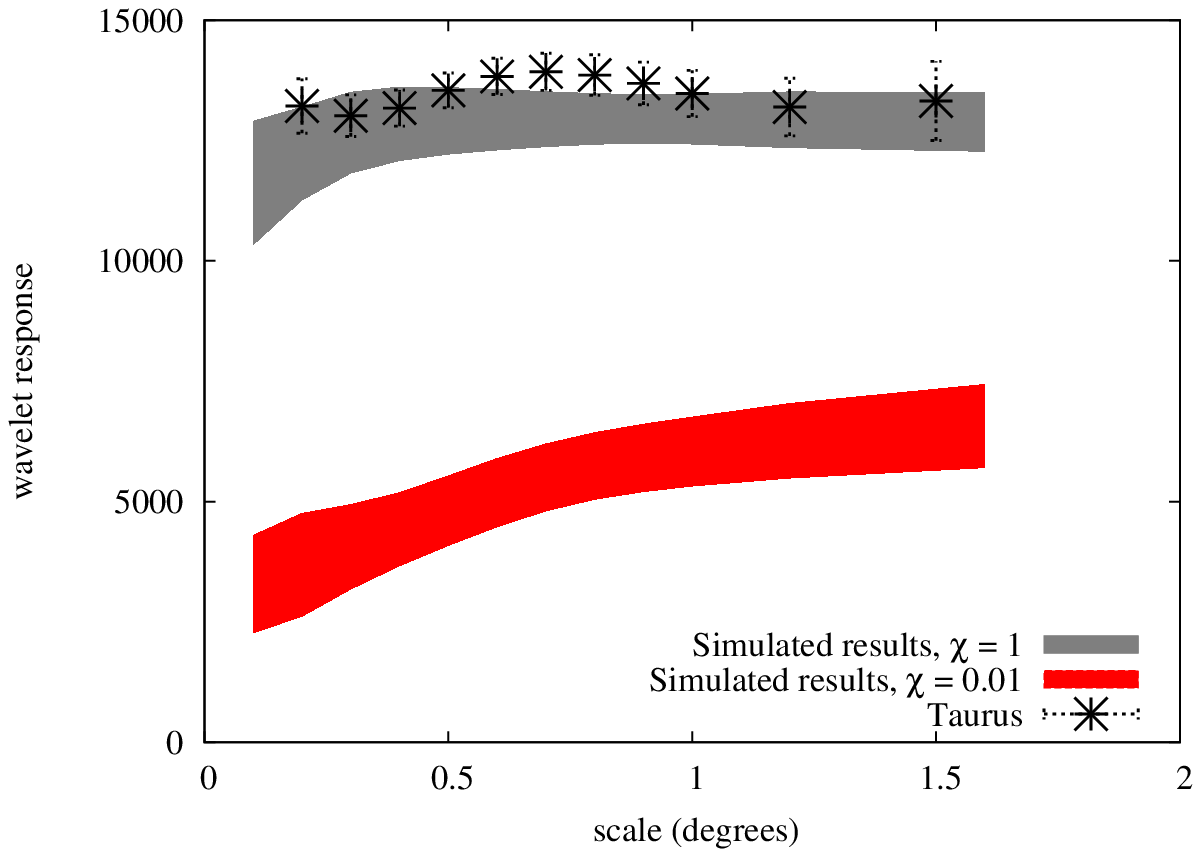}\\
\caption{The   wavelet response of Orion A  (left panel ) and Taurus (right panel), respectively. Gray shaded areas show the simulation results assuming a uniform CR density at all scales, while the red areas show the results when assuming a smaller diffusion coefficient inside MC. }
\label{fig:4}
\end{figure*}

To further investigate the relations between \gray  and dust opacity maps we also define a wavelet response as 
\begin{equation}
H_w(a)=\frac{\int \int W_1(a,{\bf x}) W^*_2(a,{\bf x})d{\bf x}}{M_1(a)},
\end{equation}
which is an analog of the impulsive response in Fourier transform. If we assume that subscript $1$ represents the dust opacity maps in Eq.3, the wavelet response should be proportional to the \gray emissivity per H-atom, that is, the CR density in different scales.  The derived  wavelet response are shown in Fig.\ref{fig:4}.  The flatness shows a uniform CR density in all scales. 

\begin{figure*}[htb]
\centering

\includegraphics[width=0.4\textwidth,angle=0]{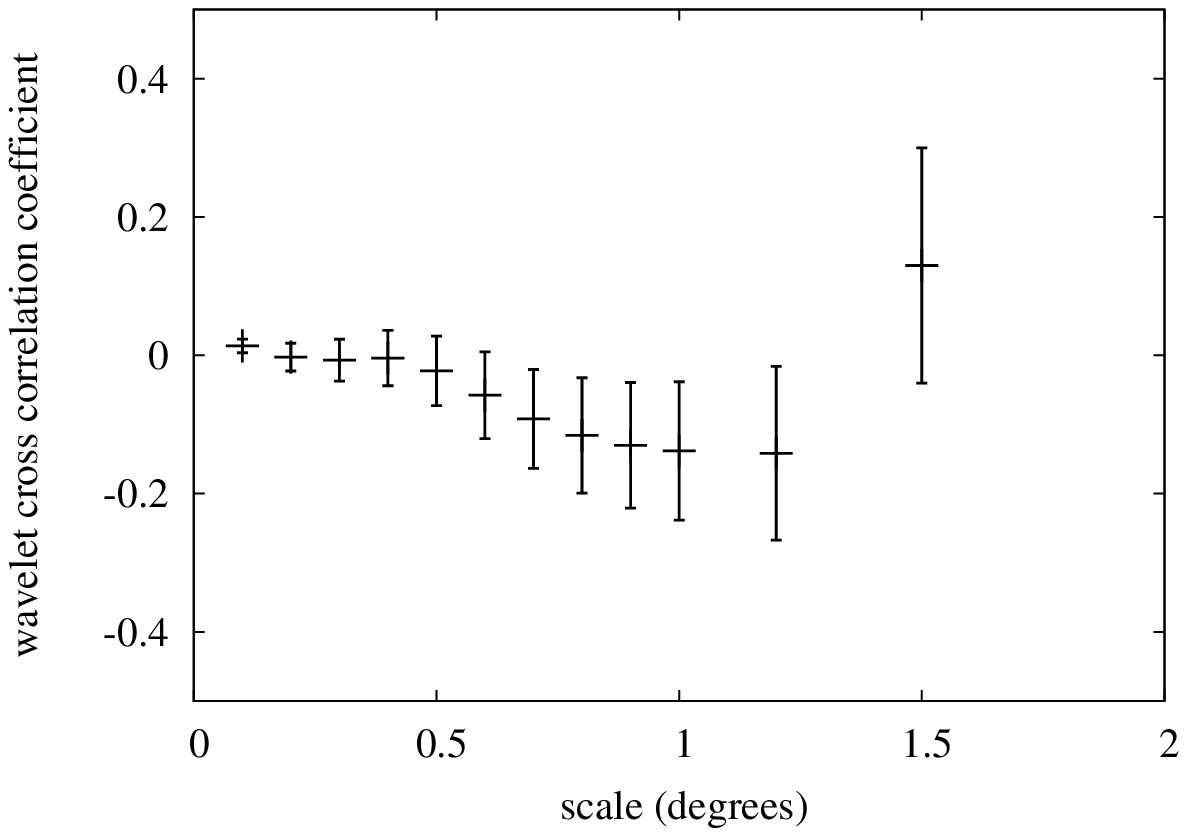}
\includegraphics[width=0.4\textwidth,angle=0]{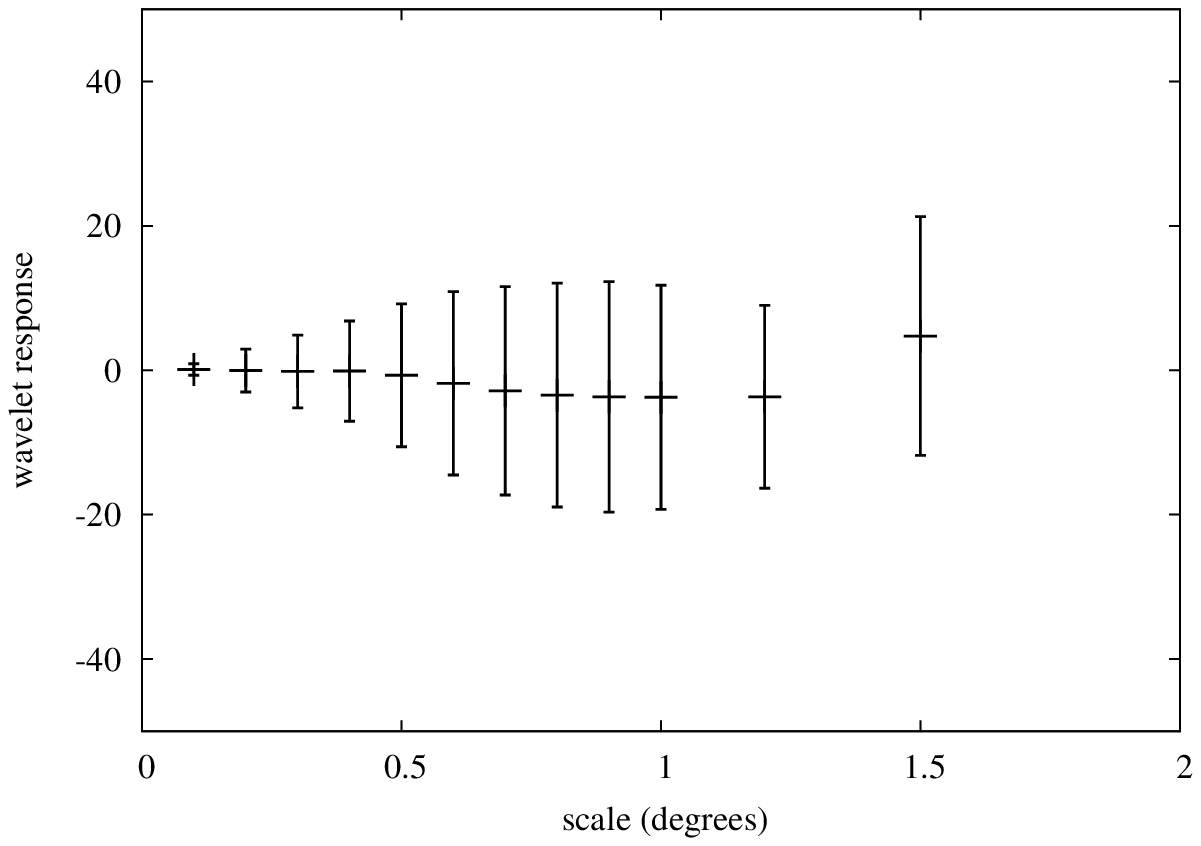}\\
\caption{Wavelet cross-correlation coefficient (left panel) and wavelet response (right panel) of 100 realization of  null test simulations. }
\label{fig:null}
\end{figure*}

To test the method we perform a null test on the random simulated maps. At each time step we sample two random maps with poisson distribution and with the same average value as the \gray map and dust column map, respectively.  Then we calculate the wavelet cross correlation coefficient and  wavelet response as described above. We perform 100 realizations of the simulations and calculate the mean and standard deviation of the wavelet cross correlation coefficient and  wavelet response. The results are shown in Fig.\ref{fig:null}, which are perfectly consistent with the zero results.

\section {CR penetrations and exclusions }
To test the power of our method we also calculate the wavelet cross correlation coefficient and wavelet response by using the simulated \gray maps, for both the CR free-penetration case and CR exclusion case. 

At first we  simulate the \fermi \gray maps by assuming a uniform CR density, which is identical to the local interstellar spectrum (LIS) of CRs \citep{casandjian_lis},  inside MC.   To do this we first calculate the \gray emissivities per Hydrogen atom  by using the LIS spectrum and the neutral pion decay cross section parametrized by \citet{kafexhiu14}. We then multiply this value with the gas column to derive the \gray emissivities at each pixel of the  $\tau_{353}$ map. We then multiply the exposure map to get the model map. Finally we  draw samples at each pixel of the "model map " from a Poisson distribution.  We then calculate the wavelet cross correlation coefficient and  wavelet response with the simulated \gray map and dust opacity maps.  The results of 100 realizations of such simulations are shown in gray shaded area in Fig. 3 and Fig. 4, which are in very good agreement with the  data. We neglect the shooting noise in the \planck opacity maps in our simulations because of the relatively higher sensitivity of \planck. 

We also simulate the results for CR exclusion cases.  The estimation of  CR exclusions from MC is done in the same framework as  described in \citet{gabici07}, where a parameter $\chi$ was adopted to describe the ratio between the diffusion coefficient inside MC and in the Galaxy.  Thus the diffusion coefficient inside MC can be parametrized as 
\begin{equation}
D(E)= 3\times10^{27} \chi(\frac{E/\rm GeV}{B/3~\rm \mu G})^{0.5}
\end{equation}
The CR inside MCs is characterized by two time scales, say, diffusion time scale $\tau_{diff}$ and cooling time scale $\tau_{cool}$, where
\begin{equation}
\tau_{diff}=\frac{R_{mc}}{6 D(E)}= 1.2\times 10^4 \chi^{-1}(\frac{R_{mc}}{20~\rm pc})^2(\frac{E}{\rm GeV})^{-0.5} (\frac{B}{10~\rm \mu G})^{0.5} yr, 
\end{equation}
and 
\begin{equation}
\tau_{cool}=\frac{1}{n_{gas} c\kappa \sigma_{pp}}= 2\times 10^5 (\frac{n_{gas}}{300~\rm cm^{-3}})^{-1} yr,.
\end{equation}
By equating $\tau_{diff}$ and $\tau_{cool}$, and taking into account the correlation
between magnetic field strength and gas density 
\begin{equation}
B=100 (\frac{n_{gas}}{10^4 ~\rm cm^-3})^{0.5},
\end{equation}
we get 
\begin{equation}
E^*=\frac{0.001}{\chi^2}(\frac{\sigma}{10^{22}~\rm cm^{-2}})^{2.5}(\frac{R_{mc}}{20~\rm pc})^{1.5}~\rm GeV,
\end{equation}
where $\sigma$ is the gas column. Below $E^*$ CR cannot penetrate into the core of MC.  It is clear for $\chi=1$ case $E^*$ is so small that all the CRs can penetrate into the cloud freely. In this case the CR density inside MC should be uniform. On the other hand, for $\chi=0.01$, $E^* \sim 10~\rm GeV$, below this value the CRs cannot penetrate the cloud  with column ${\sigma} \sim 10^{22}~\rm cm^{-2}$.  We chose  $\chi=0.01$ as our fiducial model for CR exclusion case. Remarkably,  our rule of thumb estimation came to a similar result with the numerical solution of the transport equations in \citet{gabici07} (see Fig. 1 and text therein).  Thus to calculate the \gray emissivity per H atom in this case we simply assume a sharp cutoff below $E^*$ in the CR spectrum.  Then we calculate $E^*$ for each pixel in the $\tau_{353}$ map and then calculate the corresponding \gray emissivities. The following procedure is the same as the simulations for the uniform CR case.   We note that if $E^*$ is 10 GeV the derived emissivities should be similar to the uniform CR case since in our analysis we only consider \gray maps above 2 GeV. But $E^*$ scales as $\sigma^{2.5}$ and in the ROI of Taurus and Orion A dense regions with ${\sigma} >10^{22}~\rm cm^{-2}$  exit. Thus the  \gray emissivities in the $\chi=0.01$ case should be significantly different from that in the CR uniform case. 

These results are shown in the red area in Fig. 3 and Fig. 4 . Due to the CR exclusion  the correlation between the \gray map and dust opacity map are no longer perfect, especially at smaller scales, which correspond to dense cores. In this region the CRs cannot penetrate effectively. The absolute value of wavelet response is also significantly smaller, which means a lower CR density inside MC.  The rising of wavelet response with the scales in the red shaded area in Fig.4 also illustrate the severe CRs exclusion in dense cores. 

The simulation results show a very good agreement between the data and simulations of the CR free-penetration case (uniform CR density), both in shape and absolute values. we note that the agreement in the absolute value of the the wavelet response reveal the same CR density in the MCs as the LIS. This is consistent with the analysis for the GMCs in the Gould belt in \citet{yang14}.  Thus the wavelet cross-correlation method can be used to investigate not only the CR penetration problem but also the CR densities in different structures. On the other hand, the simulation results for CR exclusion cases deviate significantly with the data and the simulations for CR free-penetration case, which show that the wavelet method can clearly distinguish these two different cases, and our result is robust.

\section{Conclusion and discussion}

The results show that the CR can penetrate the MC cores freely and the CRs density is uniform down to the scale of the \fermi angular resolution, which corresponds to 0.7 pc for Taurus and 2.6 pc for Orion A.  The current analysis was done for the \gray with energy above 2 GeV, which relate to CRs above 10 GeV.  The results reveal that there is no suppression of the diffusion coefficient in GMCs, or propagation inside the GMCs is dominated by advection. The results imply that the assumption of MC as CR barometers is correct. Furthermore, the comparison of the simulation results and data also reveal that the CR density inside the GMCs are the same as the LIS, which is consistent with the results in \citet{yang14}. 

The results show that wavelet cross correlation is a unique and powerful tool to investigate CR propagation. In the current study, this method was applied to nearby passive clouds, with \gray energy above 2 GeV. It would be also interesting to extend the analysis to lower energy to investigate  ionization loss of CRs in the MCs \citep{morlino15}. However the poor PSF of \fermi at lower energies makes the extension difficult. It would also be interesting to investigate the energy-dependent wavelet cross correlation to check weather there are energy-dependent effects of CR propagations. This would be extremely important near the young CR accelerators, for example, supernova remnants. However the Poisson noise would also limit the  application of the wavelet method. In the current study for Taurus and Orion A the Poisson noise dominates the small-scale structures for counts map above 5 GeV. For the young accelerators which are more distant than the local GMCs the statistic may be even worse. But the much larger collected areas for air Cherenkov telescope arrays (ACTA) such as HESS, MAGIC, VERITAS and the forthcoming CTA make it possible to perform this study by using the VHE image near young accelerators.

\bibliographystyle{aa}
\bibliography{ms}
\end{document}